\documentclass{ws-ijmpcs}

\begin{document}

\markboth{V.V. Anashin et al.}
{Measurement of ${\cal B}(J/\psi\to\eta_{\rm c}\gamma)$ at KEDR}

\catchline{}{}{}{}{}
%
%%%%%%%%%%%%%%%%%%%%%%%%%%%%%%%%%%%%%%%%%%%%%%%%%%%%%%%%%%%%%%%%%%%%

\title{Measurement of ${\cal B}(J/\psi\to\eta_{\rm c}\gamma)$ at KEDR\footnote{A talk presented at the 
CHARM2010 conference in Beijing, October 2010} \footnote{Partially supported by 
the Russian Foundation for Basic Research, grants 08-02-00258, 08-02-00258, and RF 
Presidential Grant for Sc. Sch. NSh-5655.2008.2.}
}

\author{
V.V. Anashin$^1$ V.M. Aulchenko$^{1,2}$ E.M. Baldin$^{1,2}$ A.K. Barladyan$^1$ A.Yu. Barnyakov$^1$ \\
M.Yu. Barnyakov$^1$ S.E. Baru$^{1,2}$ I.Yu. Basok$^1$ I.V. Bedny$^1$ O.L. Beloborodova$^{1,2}$ A.E. Blinov$^1$ \\
V.E. Blinov$^{1,3}$ A.V. Bobrov$^1$ V.S. Bobrovnikov$^1$ A.V. Bogomyagkov$^{1,2}$ A.E. Bondar$^{1,2}$ \\
A.R. Buzykaev$^1$ S.I. Eidelman$^{1,2}$ Yu.M. Glukhovchenko$^1$ V.V. Gulevich$^1$ D.V. Gusev$^1$ \\ 
S.E. Karnaev$^1$ G.V. Karpov$^1$ S.V. Karpov$^1$ T.A. Kharlamova$^{1,2}$ V.A. Kiselev$^1$ \\
S.A. Kononov$^{1,2}$ K.Yu. Kotov$^1$ E.A. Kravchenko$^{1,2}$ V.F. Kulikov$^{1,2}$ G.Ya. Kurkin$^{1,3}$ \\
E.A. Kuper$^{1,2}$ E.B. Levichev$^{1,3}$ D.A. Maksimov$^1$ V.M. Malyshev$^{1;1)}$ A.L. Maslennikov$^1$ \\
A.S. Medvedko$^{1,2}$  O.I. Meshkov$^{1,2}$ A.I. Milstein$^1$ S.I. Mishnev$^1$ I.I. Morozov$^{1,2}$ \\
N.Yu. Muchnoi$^{1,2}$  V.V. Neufeld$^1$ S.A. Nikitin$^1$ I.B. Nikolaev$^{1,2}$ I.N. Okunev$^1$ \\
A.P. Onuchin$^{1,3}$ S.B. Oreshkin$^1$ I.O. Orlov$^{1,2}$ A.A. Osipov$^1$ S.V. Peleganchuk$^1$ \\
S.G. Pivovarov$^{1,3}$ P.A. Piminov$^1$ V.V. Petrov$^1$ A.O. Poluektov$^1$ D.N. Shatilov$^1$ \\
G.E. Pospelov$^1$ V.G. Prisekin$^1$ A.A. Ruban$^1$ V.K. Sandyrev$^1$ G.A. Savinov$^1$ A.G. Shamov$^1$ \\
B.A. Shwartz$^{1,2}$ E.A. Simonov$^1$ S.V. Sinyatkin$^1$ Yu.I. Skovpen$^{1,2}$ A.N. Skrinsky$^1$ \\
V.V. Smaluk$^{1,2}$ A.V. Sokolov$^1$ A.M. Sukharev$^1$ E.V. Starostina$^{1,2}$ A.A. Talyshev$^{1,2}$ \\
V.A. Tayursky$^1$ V.I. Telnov$^{1,2}$ Yu.A. Tikhonov$^{1,2}$ K.Yu. Todyshev$^{1,2}$ G.M. Tumaikin$^1$ \\
Yu.V. Usov$^1$ A.I. Vorobiov$^1$ A.N. Yushkov$^1$ V.N. Zhilich$^1$ V.V.Zhulanov$^{1,2}$  A.N. Zhuravlev$^{1,2}$ 
}

\address{
1 Budker Institute of Nuclear Physics, 11, Lavrentiev prospect, Novosibirsk, 630090, Russia\\
2 Novosibirsk State University, 2, Pirogova street, Novosibirsk, 630090, Russia\\
3 Novosibirsk State Technical University, 20, Karl Marx prospect, Novosibirsk, 630092, Russia\\
1) E-mail: V.M.Malyshev@inp.nsk.su
}

\maketitle

\begin{abstract}
We present a study of the inclusive photon spectrum from 
6.3 million $J/\psi$ decays collected with the KEDR detector
at the VEPP-4M $e^+e^-$ collider. We measure the branching fraction 
of the radiative decay $J/\psi\to\eta_{\rm c}\gamma$, $\eta_{\rm c}$ 
width and mass. Taking into account an asymmetric  
photon line shape we obtain:
$M_{\eta_{\rm c}} = (2978.1 \pm 1.4 \pm 2.0)$ MeV/$c^2$, 
$\Gamma_{\eta_{\rm c}} = (43.5 \pm 5.4 \pm 15.8)$ MeV, 
${\cal B}(J/\psi\to\eta_{\rm c}\gamma) = (2.59\pm0.16\pm0.31)\%$.
\keywords{charmonium; radiative decays}
\end{abstract}

\ccode{PACS numbers: 13.20.Gd, 13.40.Hq, 14.40.Pq}

\section{Introduction}	

$J/\psi\to\eta_{\rm c}\gamma$ decay is a magnetic dipole radiative transition
with photon energy $\omega$ about 114 MeV, 
%, angular distribution $d\Gamma/d\Omega \sim (1+cos^2\Theta)$
and a relatively large branching fraction of about 2\%.
This is a transition between 1$S$ states of charmonium and its rate can 
be easily calculated in potential models. 
The transition does not change a spatial part of the wave function and 
its matrix element in the leading approximation equals one.
A simple calculation without relativistic corrections gives the 
result\cite{EICH} ${\cal B}(J/\psi\to\eta_{\rm c}\gamma)=3.05\%$.
It was expected that relativistic corrections are of order 20-30\%, 
similarly to the case of the electric dipole transitions 
in charmonium. 
Therefore, it was surprising when in 1986 the   
Crystal Ball Collaboration measured this branching fraction 
in the inclusive photon spectrum and obtained $(1.27\pm0.36)\%$\cite{GAIS}. 
There are a lot of theoretical predictions 
for this decay rate\cite{SHIF}$^-$\cite{EPJC}, based on QCD sum rules, 
lattice QCD calculations and so on, but as a rule they give 
values approximately twice as large as  the Crystal Ball result.

This puzzle remained for more than twenty years. Only in 2009 the CLEO
Collaboration published\cite{MITC} the result of a new measurement of this 
branching fraction using analysis of 12 exclusive decay modes of 
$\eta_{\rm c}$. The obtained value 
${\cal B}(J/\psi\to\eta_{\rm c}\gamma)=(1.98\pm0.09\pm0.30)\%$ is 
closer to theory. Combining the Crystal Ball and 
CLEO results, PDG\cite{PDG} got 
${\cal B}(J/\psi\to\eta_{\rm c}\gamma)=(1.7\pm0.4)\%$ with a scale factor 
of 1.6. In this work we report a result of the new independent measurement.

\section{Photon spectrum}

The photon spectrum in $J/\psi \to \eta_{\rm c}\gamma$ decay can be 
written as\cite{EICH}
\begin{equation}
\frac{d\Gamma(\omega)}{d\omega} = \frac{4}{3}\alpha \frac{e_{\rm c}^2}{m_{\rm c}^2}\omega^3|M|^2BW(\omega).
\end{equation}
Here $M=<\eta_{\rm c}|j_0(\omega r/2)|J/\psi>$ is the matrix element of 
the transition, 
$j_0(x)=sin(x)/x$, $e_{\rm c}$ and $m_{\rm c}$ are c-quark charge 
(in electron charge units) and mass while $BW(\omega)$ is a Breit-Wigner 
function taking into account a nonzero $\eta_{\rm c}$ width.
The matrix element tends to unity when $\omega$ tends to zero and  
decreases slowly with the photon energy increase.

CLEO found that the photon line shape of this transition 
is asymmetric, and a Breit-Wigner function alone provides a poor fit to data. 
Its modification with $\omega^3$ improves the fit around the peak, 
but gives a diverging tail at higher 
photon energies.
To suppress this tail, CLEO used in their fit 
$|M|^2=exp(-\frac{\omega^2}{8\beta^2})$ with $\beta=65$ MeV, 
explaining it by the overlap of the ground state wave functions. 
However, such a form of the matrix element is valid for the 
wave functions of the harmonic oscillator only. In all other 
potentials, $|M|^2$ dependence will be proportional to some negative degree 
of $\omega$ when $\omega$ tends to infinity. 

We tried to fit the CLEO data using another line shape: at photon energy
 $\omega$ near the resonance the decay probability 
$d\Gamma/d\omega$ is proportional to $\omega^3BW(\omega)$, 
but at higher energies the factor $\omega^3$ is replaced with $\omega$. 
We found that the function 
$d\Gamma/d\omega\sim \frac{\omega^3\omega_0^2}{\omega\omega_0+(\omega-\omega_0)^2}BW(\omega)$,
where $\omega_0=\frac{M_{J/\psi}^2-M_{\eta_{\rm c}}^2}{2M_{J/\psi}^2}$,
is also suitable. Here the correction factor 
$\frac{\omega_0^2}{\omega\omega_0+(\omega-\omega_0)^2}$ is a smooth 
 function near the resonance. 
Results of fits with the CLEO function and our function are shown 
in Table~\ref{ta1}. One can see that results on the $\eta_{\rm c}$ mass, 
width, and decay rate are close, and the confidence level of the fit with 
our function is also good. Therefore, we use the latter function 
in the analysis of our data.

\begin{table}[th]
\tbl{Results of the fits to CLEO data using various decay probability functions. 
$N_{1S}^{\rm EXC}$ is the number of signal photons in the fit.}
{\begin{tabular}{@{}ccccc@{}} \toprule
$d\Gamma/d\omega$ & $M_{\eta_{\rm c}}$, MeV/$c^2$ & 
$\Gamma_{\eta_{\rm c}}$, MeV  & $N_{1S}^{\rm EXC}$ & $\chi^2/{\rm ndf}$ (C.L.) \\ 
\colrule
$\sim\omega^3exp(-\frac{\omega^2}{8\beta^2})BW(\omega)$ & $2982.4\pm0.7$ & $32.5\pm1.8$ & $6142\pm430$ & 38.0/38 (0.47) \\
$\sim\frac{\omega^3\omega_0^2}{\omega\omega_0+(\omega-\omega_0)^2}BW(\omega)$ & $2981.8\pm0.5$ & $33.6\pm1.9$ & $6494\pm362$ & 39.1/39 (0.47) \\ \botrule
\end{tabular} \label{ta1}}
\end{table}

\section{KEDR data}
The experiment was performed at the KEDR detector\cite{KEDR} of the 
VEPP-4M collider\cite{VEPP}. 

The collider operates with a peak luminosity of about 
$1.5\cdot10^{30}$ ${\rm cm}^{-2}{\rm s}^{-1}$ near the $J/\psi$ peak energy. 
Luminosity is measured with single Bremsstrahlung online and with 
small-angle Bhabha scattering offline.
Two methods of beam energy determination are used: resonant depolarization 
with accuracy of $8\div30$ keV and IR-light Compton backscattering 
with accuracy $\sim100$ keV\cite{COMP}.

This analysis is based on a data sample of 
$1.52\pm0.08$ ${\rm pb}^{-1}$ collected at the $J/\psi$ peak.
Three $J/\psi$ scans were performed. 
Using a measured beam energy spread as well as known\cite{PDG} 
$\Gamma(J/\psi\to e^+e^-)$ and
$\Gamma(J/\psi\to hadrons)$, we calculate the $J/\psi$ production 
cross section at the peak and get 
 $N_{J/\psi}=(6.3\pm0.3)\cdot10^6$.

Event selection was performed in two steps. At the first step, 
multihadron decays of $J/\psi$ were selected with the 
following cuts: 
total energy in the calorimeters is greater than 0.8 GeV; 
at least four clusters with the energy greater than 30 MeV in 
the calorimeters are detected; 
at least one central track in the drift chamber is reconstructed; 
there are no muon tubes activated in the third layer 
of the muon system. These cuts suppress 
background from the cosmic rays, beam-gas interactions and Bhabha events. 
At the second step, photons in these events were identified. 
A photon is a cluster in the liquid krypton calorimeter which is 
not associated with the reconstructed tracks in the drift chamber
and has no time-of-flight (ToF) scintillation counters activated 
in front of it. According to the simulation, the photon detection 
efficiency with the above mentioned cuts is about 28\% and is 
almost constant in the investigated range of the photon energies.

\section{Data analysis}

\begin{figure}[ht]
\centerline{\psfig{file=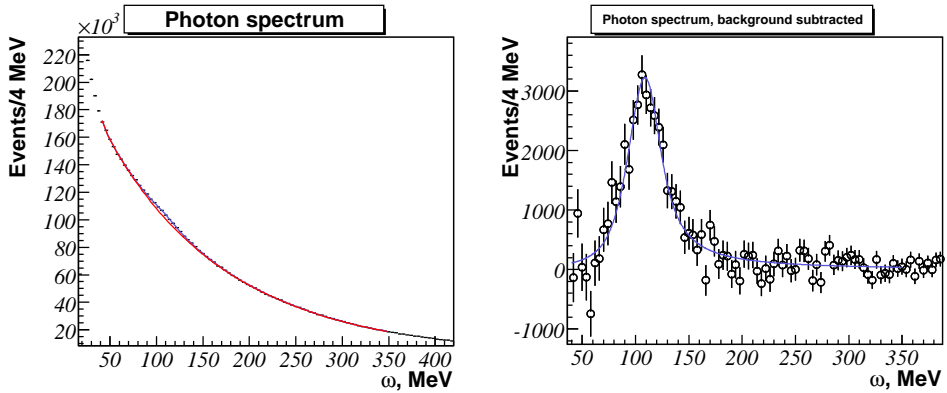,width=13.cm}}
\vspace*{8pt}
\caption{Fit of the inclusive photon spectrum. \label{fig5}}
\end{figure}

The inclusive photon spectrum and a fit to our data is shown 
in Fig.~\ref{fig5} . 
The spectrum was fit with a sum of the signal having a shape 
$d\Gamma/d\omega \sim \frac{\omega^3\omega_0^2} 
{\omega\omega_0+(\omega-\omega_0)^2} BW(\omega)$, 
convolved with the calorimeter response function 
(logarithmic normal distribution\cite{PROT} with $\sigma_E=7.4$ MeV 
at 110 MeV and a=-0.33), and background. 
The background shape was taken in the form 
$ln(dN/d\omega)=a\cdot exp(-\omega/b)+p_2(\omega)+c\cdot MIP(\omega)$, 
where the first term describes "fake" photons appearing due to  
nuclear interactions of hadrons in the calorimeter 
and usually having energies less than 60 MeV, the second term is the 
second-order polynomial describing photons arising mainly 
from $\pi^0$ decays, and $MIP(\omega)$ is the spectrum of charged 
particles (a charged particle can be also misidentified as a photon). 
The Breit-Wigner function of the form 
$BW(\omega)\sim s/((s-M^2_{\eta_{\rm c}})^2+s\Gamma^2_{\eta_{\rm c}})$, 
where $s=M^2_{J/\psi}-2\omega M_{J/\psi}$, was used in the fit.

We also tried to fit our data using other line shapes. We used 
$BW(\omega)$ alone, $\omega^3 BW(\omega) $,
the CLEO function, and our function. 
Again, as first noted by CLEO, we found that the $BW(\omega)$ function 
alone gives a shifted value of $\eta_{\rm c}$ 
mass compared to other functions: $M(\eta_{\rm c})=2974.3\pm1.4$ MeV/$c^2$. 
The function $\omega^3BW(\omega)$ leads to a large tail at higher 
photon energies, giving for the branching fraction
${\cal B}(J/\psi\to\eta_{\rm c}\gamma) = (7.3\pm0.5)\%$ 
(here the decay probability function was integrated up to $M_{J/\psi}/2$).
The last two functions give close fit results. Since the CLEO function
has an exponential factor, the result for the branching fraction
with this function can be considered as a lower limit. The difference 
between the results obtained with the two last functions is  used to
estimate a systematic error appearing due to the unknown line shape.

\section{Systematic errors}
Systematic errors of our measurements are listed in Table~\ref{ta3}. 

\begin{table}[ht]
\tbl{ Systematic errors.}
{\begin{tabular}{@{}lccc@{}} \toprule
 Systematic error & $M_{\eta_{\rm c}}$, MeV/$c^2$ & 
$\Gamma_{\eta_{\rm c}}$, MeV  & ${\cal B}(J/\psi)\to\gamma\eta_{\rm c}), \%$ \\
\colrule
 Line shape & 0.7 & 2.3 & 0.15  \\
 $\eta_{\rm c}$ width & 0.4 &  & 0.15  \\
 Background subtraction & 0.8 & 15.6 & 0.17  \\
 Number of $J/\psi$ produced &  &  & 0.13  \\
 Photon efficiency &  &  & 0.08  \\
 Photon energy scale & 1.7 &  &   \\
\colrule
 Total & 2.0 & 15.8 & 0.31  \\
\botrule
\end{tabular} \label{ta3}}
\end{table}

The uncertainty of the $\eta_{\rm c}$ width 
leads to an error, which we evaluated varying the 
$\eta_{\rm c}$ width in the fit by 2.2 MeV (the current PDG error).
%leads to an error, which we evaluated by fixing the 
%$\eta_{\rm c}$ width to the current PDG values ($28.6\pm2.2$ MeV).
A systematic error related to the background subtraction was estimated 
by using in the fit a polynomial of the third order 
instead of the second-order one, varying ranges of the fit, 
and applying or not the ToF veto in photon selection.
The error in the number of $J/\psi$ produced was evaluated using the
known uncertainty of the luminosity measurement.
Since the cluster multiplicity is different in the simulation and experimental 
photon spectrum in $J/\psi$ decays, the error due to the
photon detection efficiency was estimated by changing by 25\% weights of events 
with small $(n<4)$ and large
 $(n>3)$ track multiplicities, and taking different MC generators 
for the $\eta_{\rm c}$ decays.
The calibration of the photon energy scale  was performed using 
$\pi^0\to 2\gamma$ decays and 
$\psi'\to\gamma\chi_{cJ}\to\gamma J/\psi$ transitions.

\section{Results and conclusions}
A new direct measurement of $J/\psi\to\eta_{\rm c}\gamma$ decay was performed. 
We measured the $\eta_{\rm c}$ mass, width, and branching fraction 
of $J/\psi\to\eta_{\rm c}\gamma$ decay. 
The values of the branching fraction and $\eta_{\rm c}$ mass are 
sensitive to the line shape of the photon spectrum 
and it should be taken into account during analysis. 
Our results on the $\eta_{\rm c}$ mass and width are: 

\begin{center}
$M_{\eta_{\rm c}} = 2978.1 \pm 1.4 \pm 2.0$ MeV/$c^2$,

$\Gamma_{\eta_{\rm c}} = 43.5 \pm 5.4 \pm 15.8$ MeV.

\end{center}

Before our experiment these parameters were measured in $J/\psi$ and 
$B$ meson decays, 
as well as in $\gamma\gamma$ and $p\bar p$ collisions. 
Measurements of Crystal Ball, MARK3, BES, and KEDR were performed 
using the radiative $J/\psi$ decays, therefore, 
a mass shift should be taken into account. Crystal Ball and KEDR made 
such a correction, but MARK3 and BES did not. 
Therefore  we believe that MARK3 and BES results on the $\eta_{\rm c}$ 
mass should be corrected by approximately 4 MeV 
towards higher values.

Our result on the branching fraction of $J/\psi\to\eta_{\rm c}\gamma$ decay is
\begin{center}
${\cal B}(J/\psi\to\eta_{\rm c}\gamma) = (2.59\pm0.16\pm0.31)\%$.
\end{center}
It is consistent with that of CLEO, is higher than 
the old Crystal Ball value and close to theoretical predictions. 
%Although the difference between our result and 
%the Crystal Ball one is sizable, we think  that it can appear due to the 
%fact that Crystal Ball ascribes a statistical error to their result only 
%assuming that a systematic one is negligible. 

The authors are grateful to N. Brambilla and A. Vairo 
for useful discussions.

\end{document}